\documentclass[apj]{emulateapj}

\newcommand{\Te} {T_{\rm eff}}
\newcommand{\logg} {\log g} 
\shorttitle{6 NEW ZZ CETI WHITE DWARFS}
\shortauthors{GIANNINAS ET AL.}
\slugcomment{Accepted for Publication in the Astronomical Journal}
\begin{document}


\title{MAPPING THE ZZ CETI INSTABILITY STRIP: DISCOVERY OF 6 NEW PULSATORS}

\author{A. Gianninas, P. Bergeron, and G. Fontaine}
\affil{D\'epartement de Physique, Universit\'e de Montr\'eal, C.P.~6128, 
Succ.~Centre-Ville, Montr\'eal, Qu\'ebec, Canada, H3C 3J7.}
\email{gianninas@astro.umontreal.ca, fontaine@astro.umontreal.ca, 
bergeron@astro.umontreal.ca}

\begin{abstract}
As part of an ongoing program to map better the empirical instability
strip of pulsating ZZ Ceti white dwarfs, we present a brief progress
report based on our last observing season. We discuss here high-speed
photometric measurements for 6 new pulsators. These stars were
selected on the basis of preliminary measurements of their effective
temperature and surface gravity that placed them inside or near the
known ZZ Ceti instability strip. We also report detection limits for a
number of DA white dwarfs that showed no sign of variability. Finally,
we revisit the ZZ Ceti star G232-38 for which we obtained improved
high-speed photometry.
\end{abstract}


\section{MOTIVATION}

The ZZ Ceti stars are pulsating white dwarfs whose optical spectra are
dominated by hydrogen lines (DA stars). They are found in a very well
defined region in the $\Te$-$\logg$ plane, known as the ZZ Ceti
instability strip. They exhibit pulsation periods in the range 100-1400
s, corresponding to low degree gravity-mode oscillations. The detected
pulsation modes have amplitudes that are found in the range from a few
millimagnitudes in the low-amplitude pulsators to a fraction of a
magnitude in the largest amplitude ones. The ZZ Ceti stars show some 
remarkable trends, although there are a few exceptions. For instance,
the cooler pulsators tend to show longer periods and larger amplitudes. 
Also, for a given effective temperature, the observed periods tend to be
longer for lower gravity objects.

Since the early 90's, our group in Montr\'eal has been carrying out a
systematic study aimed at defining the empirical boundaries of the ZZ
Ceti instability strip. We use quantitative time-averaged optical
spectroscopy to pin down the locations of candidate stars in the
$\Te$-$\logg$ diagram, and we follow up on these candidates with
``white light'' fast photometry to determine if a target pulsates or
not. The determination of these boundaries is essential if we are to
understand better the ZZ Ceti phenomenon and to establish on a more
quantitative footing the trends already alluded to above. The question
of the purity of the instability strip (i.e., the absence of
photometrically constant stars within) is also an important issue, and
this for several reasons. First, a pure strip implies that the ZZ Ceti
phenomenon is an evolutionary phase through which all DA white dwarfs
must pass. It is then possible to apply what is gleaned from
asteroseismological studies of these stars, such as information on
their internal structure, to the entire class of DA white
dwarfs. Second, knowing that the strip is pure allows us to predict
the variability of a star simply through a measure of its atmospheric
parameters, $\Te$ and $\logg$.

The paper of \citet{bergeron95} presented, for the first time, a
complete and homogeneous picture of the ZZ Ceti stars in terms of
their time-averaged properties. It was based on the sample of 22
pulsators known at the time. In the meantime, we pursued our survey to
improve statistics, and we reported several discoveries and new
findings in a series of papers, the last of which is
\citet{gianninas05} where references to our previous work can be
found. One of the key results of \citet{gianninas05}, based on our
homogeneous spectroscopic sample of 39 pulsators and 121 constant
stars, is the demonstration that the ZZ Ceti strip is indeed
pure. Furthermore, that paper incorporated the first results of an
extended ongoing survey whose primary aim is to define more accurately
the empirical boundaries of the instability strip. That survey is
based on spectroscopic measurements of a subsample of DA white dwarfs
from the Catalog of Spectroscopically Identified White Dwarfs of
\citet{mccook99}. We thus obtain optical spectra for each star and
then fit the Balmer line profiles using a grid of synthetic spectra
generated from detailed model atmospheres. This allows us to
accurately measure $\Te$ and $\logg$ for each star. It is then
possible to identify white dwarfs whose atmospheric parameters place
them near or within the ZZ Ceti instability strip. In this paper, we
report on the latest results of this ongoing survey after the
2005-2006 observing season.

\begin{table*}[!t]
\caption{Journal of Observations}
\begin{center}
\begin{tabular}{clc@{}cccc}
\hline
\hline
   &      & \multicolumn{2}{c}{Run}    & Date & Start Time & Total Number of \\
WD & Name & \multicolumn{2}{c}{Number} &  UT  &     UT     &   Data Points   \\
\hline
0016$-$258 & MCT 0016$-$2553 & kit-013 & & 2005 Oct 27 & 05:43 &  867 \\
0036+312   & G132-12         & kit-005 & & 2005 Oct 25 & 05:25 & 1042 \\
           &                 & kit-017 & & 2005 Oct 28 & 04:42 & 1348 \\
0136+768   & GD 420          & kit-009 & & 2005 Oct 26 & 05:55 &  410 \\
0145+234   & MK 362          & kit-002 & & 2005 Oct 24 & 07:28 &  736 \\
0302+621   & GD 426          & kit-010 & & 2005 Oct 26 & 07:18 &  532 \\
0344+073   & KUV 03442+0719  & kit-003 & & 2005 Oct 24 & 10:04 &  899 \\
           &                 & kit-011 & & 2005 Oct 26 & 09:07 & 1188 \\
           &                 & kit-015 & & 2005 Oct 27 & 11:25 &  366 \\
0347$-$137 & GD 51           & kit-006 & & 2005 Oct 25 & 08:49 &  505 \\
0416+701   & GD 429          & kit-014 & & 2005 Oct 27 & 08:54 &  801 \\
0741+248   & LP 366-3        & kit-007 & & 2005 Oct 25 & 10:31 &  737 \\
1150$-$153 & EC 11507$-$1519 & kit-021 & & 2006 Mar 23 & 08:47 &  723 \\
1211+320   & CBS 54          & kit-018 & & 2006 Mar 23 & 03:43 &  582 \\
1503$-$093 & EC 15036$-$0918 & kit-022 & & 2006 Mar 23 & 11:11 &  439 \\
1820+709   & GD 530          & cfh-111 & & 2005 May 01 & 13:42 &  328 \\
           &                 & cfh-119 & $^{\rm a}$ & 2005 Aug 14 & 05:52 &  468 \\
           &                 & kit-012 & & 2005 Oct 27 & 02:21 & 1048 \\
           &                 & kit-016 & & 2005 Oct 28 & 01:54 &  918 \\
2148+539   & G232-38         & kit-004 & & 2005 Oct 25 & 02:04 & 1082 \\
2148$-$291 & MCT 2148$-$2911 & cfh-117 & $^{\rm a}$ & 2005 Aug 13 & 07:33 &  382 \\
2311+552   & GD 556          & kit-001 & & 2005 Oct 24 & 02:17 & 1724 \\
2336$-$079 & GD 1212         & kit-008 & & 2005 Oct 26 & 02:23 & 1083 \\
\hline
\multicolumn{7}{l}{$^{\rm a}$ Observed through a B-band filter.} \\
\multicolumn{7}{l}{Notes. -- The sampling time is 10 s for all observations.} \\
\end{tabular}
\end{center}
\end{table*}

We note that with the recent releases of data sets from the SDSS, the
number of known ZZ Ceti stars has grown in a spectacular way, by more
than a factor of 2.5 \citep[see,
e.g.,][]{mukadam04b,mullally05,kepler05,castanheira06}. The vast
majority of the new pulsators, however, are significantly fainter than
those previously known and, therefore, have a much less interesting
asteroseismological potential. Moreover, the relatively low S/N ratio
spectroscopic observations reached in the SDSS limits the usefulness
of statistical studies based on them. Of particular interest here,
\citet{mukadam04a} claimed that the ZZ Ceti instability strip contains
several nonvariable stars on the basis of the SDSS sample of
\citet{mukadam04b}. We conclusively showed in \citet{gianninas05},
however, that this incorrect result could be explained in terms of
insufficient S/N ratio in the SDSS spectra. Thus, we believe that
improvements in our empirical description of the ZZ Ceti instability
strip necessarily rests on high S/N ($>$ 80) time-averaged
spectroscopy such as in the survey we are currently pursuing. This
limits the potential targets to relatively bright stars ($V < 17$),
but as we are finding out as well as others \citep[see,
e.g.,][]{silvotti05,voss06}, there are still bright ZZ Ceti pulsators
to be discovered.

\section{RESULTS}

\begin{figure}[!t]
\plotone{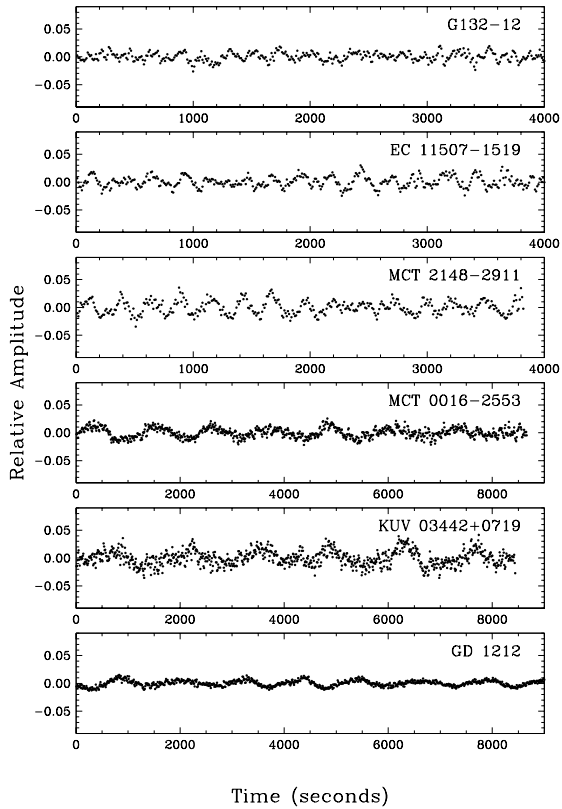}
\caption{Light curves of the 6 new ZZ Ceti stars observed in ``white''
light by LAPOUNE at Steward Observatory's 2.3 m telescope with the
exception of MCT 2148$-$2911 that was observed with LAPOUNE attached
to the 3.6 m Canada-France-Hawaii telescope through a B-band
filter. Each point represents a sampling time of 10 s. The light
curves are expressed in terms of residual amplitude relative to the
mean brightness of the star.}
\label{lcurves}
\end{figure}

\begin{figure}[!t]
\plotone{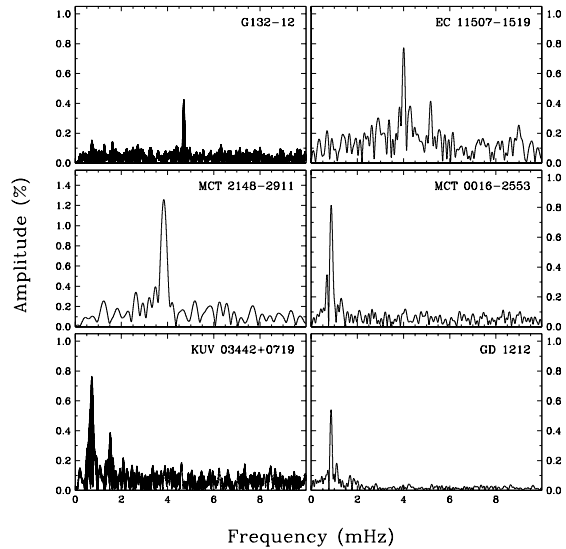}
\caption{Fourier (amplitude) spectra for the light curves of the 6 new
ZZ Ceti stars in the 0-10 mHz bandpass. The spectrum in the region
from 10 mHz to the Nyquist frequency is entirely consistent with noise
and is not shown. The amplitude scales are identical for each star for
the purpose of comparison with the exception of MCT 2148$-$2911 due to
its larger amplitude. The amplitude axis is expressed in terms of the
percentage variations about the mean brightness of the star.}
\label{ffts}
\end{figure}

We proceeded to secure high-speed photometric measurements for those
newly analyzed stars in our spectroscopic survey whose atmospheric
parameters would place them inside or near the edges (blue and red) of
the known ZZ Ceti instability domain. These follow-up observations
were secured over the course of 3 observing runs in 2005 and 2006. In
all, 6 of these stars turned out to be genuine pulsating DA white
dwarfs: ZZ Ceti stars. A summary of our observations is presented in
Table 1. MCT 2148$-$2911 was observed on 2005 August 13 with the 3.6
m Canada-France-Hawaii telescope equipped with LAPOUNE, the portable
Montr\'eal three-channel photometer. MCT 0016$-$2553, G132-12, KUV
03442+0719, and GD 1212 were observed during a 5 night run from 2005
October 23 to 27 at the Steward Observatory 2.3 m telescope, equipped
once again with LAPOUNE. Finally, EC 11507$-$1519 was observed on 2006
March 22 at Steward Observatory with the same telescope and
setup. Figure \ref{lcurves} displays the sky-subtracted,
extinction-corrected light curves obtained for each star. The
resulting Fourier (amplitude) spectra are shown in Figure \ref{ffts}.

Summarized in Table 2 are the data for the 6 new ZZ Ceti stars split
into short and long period variables. We list the $V$ magnitude as
well as the amplitude and period of the dominant pulsation mode
obtained from the corresponding Fourier spectra in Figure
\ref{ffts}. In Table 3, we provide the data for the white dwarfs we found to be
photometrically constant. In addition to the apparent magnitude, we
also indicate the limit to which no photometric variations were
detected (in the 20-2000 s period range). This limit corresponds to
three times the mean noise level of the Fourier transform in that
period range. It is expressed as a percentage of the mean brightness
of the star.

\begin{table}
\caption{Photometric parameters of new ZZ Ceti white dwarfs}
\begin{center}
\begin{tabular}{clccc}
\hline
\hline
   &       &         & Amplitude & Dominant   \\
WD &  Name & {\it V} &   (\%)    & Period (s) \\
\hline
\multicolumn{5}{c}{Short Period}\\
\hline
0036+312   & G132-12         & 16.20 & 0.43 &  212.7 \\
1150$-$153 & EC 11507$-$1519 & 16.00 & 0.77 &  249.6 \\
2148$-$291 & MCT 2148$-$2911 & 16.10 & 1.26 &  260.8 \\
\hline
\multicolumn{5}{c}{Long Period}\\
\hline
0016$-$258 & MCT 0016$-$2553 & 16.10 & 0.81 & 1152.4 \\
0344+073   & KUV 03442+0719  & 16.10 & 0.76 & 1384.9 \\
2336$-$079 & GD 1212         & 13.26 & 0.54 & 1160.7 \\
\hline
\end{tabular}
\end{center}
\end{table}

The atmospheric parameters for the 6 new ZZ Ceti stars and the 10
photometrically constant DA white dwarfs are reported in Table 4
together with the stellar masses and absolute visual magnitudes. Our
theoretical framework and fitting technique are described at length in
\citet{gianninas05} and references therein. To obtain proper time-averaged 
spectra for ZZ Ceti stars, it is necessary to set the exposure time
long enough to cover several pulsation cycles. However, not knowing a
priori that the stars listed in Table 2 would turn out to be variable,
this criterion was not met for the three longer period variables. It
will therefore be necessary to acquire new optical spectra for these
stars in order to refine our determination of their atmospheric
parameters. Consequently, the atmospheric parameters for these objects
are considered preliminary and as such, they are marked with a colon in Table 4.
Nonetheless, these results demonstrate once again the ability of the
spectroscopic method, pioneered by \citet{bergeron95}, at predicting
the variability of DA white dwarfs.

\begin{table}[!h]
\caption{Photometrically Constant DA White Dwarfs}
\begin{center}
\begin{tabular}{clc@{}ccc}
\hline
\hline
   &        & \multicolumn{2}{c}{}      & Detection limit \\
WD &  Name  & \multicolumn{2}{c}{\it V} &      (\%)       \\  
\hline
0136+768   & GD 420          & 14.85 &           & 0.13 \\
0145+234   & MK 362          & 14.50 &$^{\rm a}$ & 0.06 \\
0302+621   & GD 426          & 14.95 &           & 0.19 \\
0347$-$137 & GD 51           & 14.00 &$^{\rm a}$ & 0.15 \\
0416+701   & GD 429          & 14.74 &           & 0.07 \\
0741+248   & LP 366-3        & 16.52 &           & 0.34 \\
1211+320   & CBS 54          & 16.00 &$^{\rm a}$ & 0.54 \\
1503$-$093 & EC 15036$-$0918 & 15.15 &           & 0.27 \\
1820+709   & GD 530          & 17.00 &           & 0.56 \\
2311+552   & GD 556          & 16.17 &           & 0.15 \\
\hline
\multicolumn{5}{l}{$^{\rm a}$Photographic B magnitudes.}
\end{tabular}
\end{center}
\end{table}

\begin{figure}[!h]
\plotone{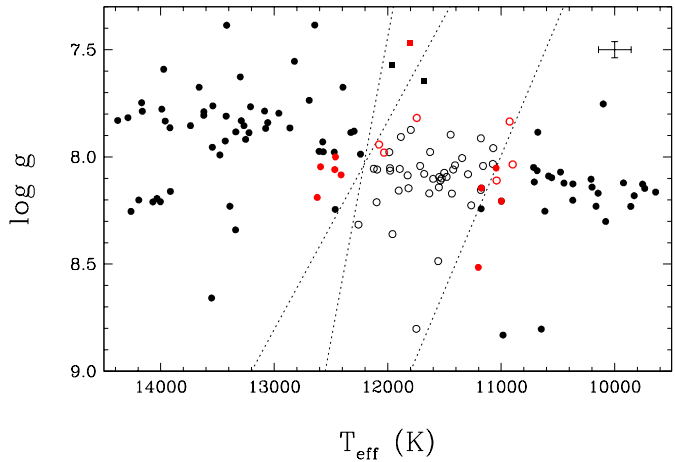}
\caption{$\Te-\logg$ distribution for DA white dwarfs with high-speed
photometric measurements. The black open and filled circles represent
respectively the 39 ZZ Ceti stars and the 121 constant DA stars taken
from \citet[][and references therein]{gianninas05}, while the red open
and filled circles represent the 6 new ZZ Ceti stars and the 10
constant DA stars reported in Tables 2 and 3, respectively.  The
filled squares correspond to unresolved double degenerate systems (see
text).  The error bars in the upper right corner represent the average
uncertainties of the spectroscopic method in the region of the ZZ Ceti
stars (1.2\% in $\Te$ and 0.038 dex in $\logg$). The dashed lines
represent the empirical blue and red edges of the instability strip
allowed by our current atmospheric parameter determinations for these
stars.}
\label{zzstrip}
\end{figure}

Our updated ZZ Ceti instability strip is displayed in Figure
\ref{zzstrip} together with the empirical blue and red edges allowed by
our current atmospheric parameter determinations for these stars.
Note that three low-gravity objects, shown by filled squares in Figure
\ref{zzstrip}, represent unresolved double degenerate systems and that
the atmospheric parameters
 are the average values of both components
of the system; the red
 square is GD 429 (WD~0416+701) identified as a
velocity variable by
 \citet{maxted00}. Hence these
 cannot be used to
constrain the slope of the blue edge, as discussed by
\citet{gianninas05}. Furthermore, since the atmospheric parameters for
the new 
 long period variables are preliminary, we have refrained
from
 redefining the empirical red edge, and we have simply reproduced
in
 Figure \ref{zzstrip} the value obtained by \citet{gianninas05}.
It is
 clear that additional observations are required to pin down the
final
 empirical boundaries, in particular near the blue edge of the
strip.
 Hopefully we will report such observations in the near future
when our
 survey is completed.
 
 We note that the periods of the
dominant oscillation modes in the
 first three stars of Table 2, as
well as those of the last three
 objects, are consistent with the
positions of these stars inside or
 near the blue (short-period
pulsators) and red (long-period pulsators)
 edges of the instability
strip, respectively. However, as a curiosity,
 we also point out that
the amplitudes of the detected modes in the
 three ``red edge''
pulsators are not very large, especially for GD
 1212. This goes
against the general tendency to observe large
 amplitudes in the
cooler objects. Note also the very long period seen
 in KUV
03442+0719, one of the longest if not the longest one ever
 detected
in a ZZ Ceti star. Are we observing stars at the very red
 edge of the
strip, basically ``dying off'' in amplitude?

\begin{table}[!h]
\caption{Atmospheric Parameters of New ZZ Ceti and Photometrically Constant DA Stars}
\begin{center}
\begin{tabular}{clcccc}
\hline
\hline
   &      & $T_{\rm eff}$ &          &               &       \\
WD & Name &      (K)      & $\log g$ & $M/M_{\odot}$ & $M_V$ \\
\hline
\multicolumn{6}{c}{ZZ Ceti} \\
\hline
0016$-$258 & MCT 0016$-$2553 & 10900: & 8.04: & 0.63 & 11.92 \\
0036+312   & G132-12         & 12080  & 7.94  & 0.57 & 11.53 \\
0344+073   & KUV 03442+0719  & 10930: & 7.84: & 0.51 & 11.62 \\
1150$-$153 & EC 11507$-$1519 & 12030  & 7.98  & 0.60 & 11.59 \\
2148$-$291 & MCT 2148$-$2911 & 11740  & 7.82  & 0.51 & 11.42 \\
2336$-$079 & GD 1212         & 11040: & 8.11: & 0.67 & 11.99 \\
\hline
\multicolumn{6}{c}{Photometrically Constant} \\
\hline
0136+768   & GD 420          & 11050  & 8.05  & 0.64 & 11.90 \\
0145+234   & MK 362          & 12470  & 8.06  & 0.64 & 11.64 \\
0302+621   & GD 426          & 11000  & 8.21  & 0.73 & 12.15 \\
0347$-$137 & GD 51           & 12620  & 8.19  & 0.72 & 11.81 \\
0416+701   & GD 429          & 11810  & 7.47  & 0.35 & 10.94 \\
0741+248   & LP 366-3        & 12410  & 8.08  & 0.66 & 11.69 \\
1211+320   & CBS 54          & 12460  & 8.00  & 0.61 & 11.56 \\
1503$-$093 & EC 15036$-$0918 & 12590  & 8.05  & 0.64 & 11.61 \\
1820+709   & GD 530          & 11200  & 8.52  & 0.94 & 12.61 \\
2311+552   & GD 556          & 11180  & 8.15  & 0.69 & 12.01 \\
\hline
\end{tabular}
\end{center}
\end{table}

\begin{figure}[h]
\plotone{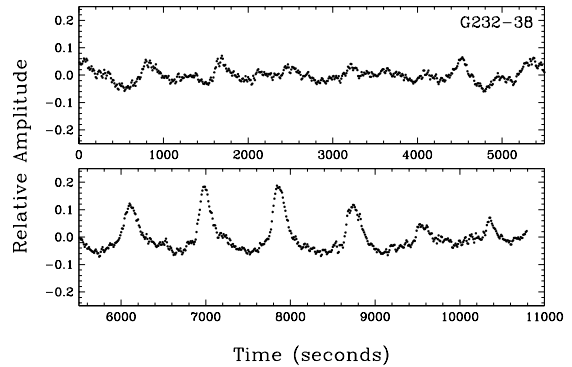}
\caption{Same as Fig. \ref{lcurves}, but for G232-38.}
\label{lcurve_g232}
\end{figure}

Finally, during the 2005 October observation run, the observing
conditions were such that we were also able to revisit G232-38 whose
discovery as a ZZ Ceti star was reported in Gianninas et al. (2005).
Indeed, the discovery light curve \citep[see Fig.~2 of][]{gianninas05}
had been obtained with LAPOUNE attached to the 1.6 m telescope of the
Observatoire du mont M\'egantic under less than ideal
conditions. Having a 2.3 m telescope at our disposal, we were certain
we could achieve a better result. The new light curve we obtained,
which is just over 3 hrs long (see Table 1), is shown in Figure
\ref{lcurve_g232}. 

\begin{figure}
\plotone{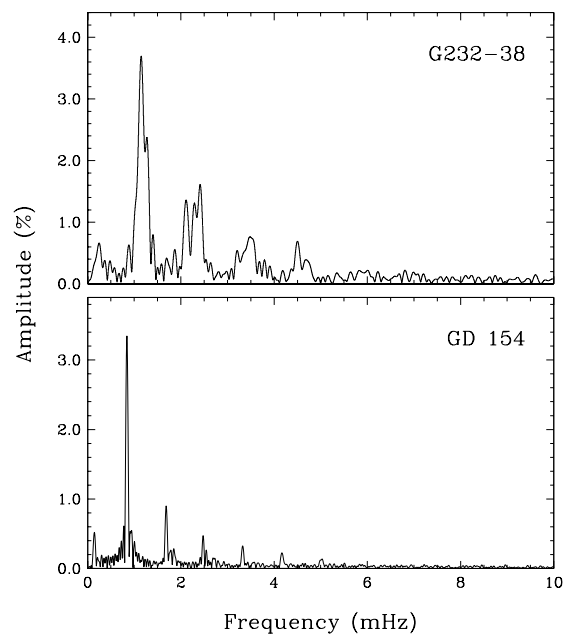}
\caption{Same as Fig. \ref{ffts}, but for G232-38 ({\it top}) and GD 154 ({\it bottom}).}
\label{fft_g232}
\end{figure}

\noindent On inspection, we see that this new observation is 
indeed much better than the previous one with the pulsations very
clearly visible. The amplitude modulation due to destructive
interference between the different pulsation modes is also very
evident. From this light curve, we computed the Fourier (amplitude) 
spectrum shown in the top panel of Figure \ref{fft_g232}. The dominant
pulsation period obtained from the Fourier spectrum is 870.4 s. The
light curve and associated Fourier spectrum are, in fact, quite
reminiscent of those found in another cool large amplitude ZZ Ceti
star, GD 154 \citep{pfeiffer96} whose Fourier spectrum is shown in the
bottom panel of Figure~\ref{fft_g232}. The Fourier spectrum of GD 154
was calculated from a nearly 6-hour long light curve which we obtained
on the night of 1991 May 22 with LAPOUNE attached to the 3.6 m
Canada-France-Hawaii telescope. It is likely, as in GD 154, that the
main peak in the Fourier spectrum of G232-38 contains a few closely
spaced modes in frequency, and that the other peaks are simply
nonlinear structures caused by harmonics and cross-frequencies of the
main modes.

\acknowledgements

We would like to thank the director and staff of Steward Observatory
and the Canada-France-Hawaii Telescope for the use of their facilities
and for supporting LAPOUNE as a visitor instrument. This work was
supported in part by the NSERC Canada. A. G. acknowledges the
contribution of the Canadian Graduate Scholarships. G. F. acknowledges
the contribution of the Canada Research Chair Program. P. B. is a
Cottrell Scholar of Research Corporation.

\end{document}